\documentclass[12pt]{article}
\usepackage{amssymb}

\newcommand{\be}{\begin{equation}}
\newcommand{\ee}{\end{equation}}
\def\bea{\begin{eqnarray}}
\def\eea{\end{eqnarray}}

\newcommand{\bn}{\begin{eqnarray}}
\newcommand{\en}{\end{eqnarray}}

\newcommand{\p}{\partial}

\newcommand{\nn}{\nonumber}

\newcommand{\no}{\noindent}

\newcommand{\tb}{\tilde{B}}

\def\bea{\begin{eqnarray}}
\def\eea{\end{eqnarray}}

\newcommand{\beq}{\begin{eqnarray}}
\newcommand{\eeq}{\end{eqnarray}}
\newcommand{\omw}{\mathbb{W}^{TT}}

\begin{document}

\title{\textbf{More on the spin-2 analogue of the massive BF model}}
\author{{ A. L. R. dos Santos\footnote{alessandroluiz1905@gmail.com}, D. Dalmazi\footnote{denis.dalmazi@unesp.br} }\\
\textit{{UNESP - Campus de Guaratinguet\'a - DFI} }\\
\textit{{CEP 12516-410, Guaratinguet\'a - SP - Brazil.} }\\}
\date{\today}
\maketitle

\begin{abstract}

The addition of mass terms in general breaks gauge symmetries  which can  be recovered usually via Stueckelberg fields. The massive BF model describes massive spin-1 particles  while preserving the $U(1)$ symmetry without Stueckelberg fields. Replacing the spin-1 curvature (field strength) by  the Riemann tensor one can define its spin-2 analogue (massive``BR'' model). Here we investigate the canonical structure of the free mBR  model in terms of gauge invariants in arbitrary dimensions and compare with the massive BF model.  We also investigate  non linear completions of the mBR  model in arbitrary dimensions. In $D=3$ we find a non linear completion in the form of a bimetric model which is a  sub case of  a new class of bimetric models whose decoupling limit is ghost free at leading order. Their spectrum consists only of massive spin-2 particles. In arbitrary dimensions $D\ge 3$  we show that the consistency of a possible single metric completion of the mBR  model is related with the consistency  of a higher rank description of massless spin-1 particles in arbitrary backgrounds.

\end{abstract}

\newpage

\section{Introduction}

The universal nature of the gravitational interaction makes the search for a possible graviton mass a fundamental subject. Severe problems in the consistency of massive gravitons like the vDVZ mass discontinuity \cite{vDV,Zak}, and the presence of ghosts \cite{bd} have been tackled by a convenient choice of the graviton potential \cite{deRham-2}. Such work has triggered a huge amount of work in the subject of massive gravity, see \cite{Hinter-1,deRham-1}. In particular, the need of viable cosmological solutions has led to the bimetric model of \cite{hr} which besides a massive spin-2 particle contains an extra massless spin-2 particle. Regarding the phenomenology and consistency of \cite{hr}, see \cite{hogas1,hogas2}.

The model \cite{deRham-2} and also \cite{hr} are based on the  Fierz-Pauli massive spin-2 free theory \cite{Fierz} which can be defined in terms of a symmetric rank-2 tensor. One might wonder how robust are the physical outputs of the present massive gravity theories against the replacement of the FP paradigm by other descriptions of free massive spin-2 particles.

The introduction of mass terms breaks the local symmetries of the massless theories in general. In the case of the metric formulation of gravity the local symmetry is reparametrization invariance which should be preserved also in the massive theory. This is achieved in  \cite{hr} via the introduction of a second metric which is a dynamical field and enlarges the spectrum of the theory as compared to \cite{deRham-2}. One migh search for a phenomenologically viable massive gravity with only massive spin-2 particles in the spectrum this is the main motivation for the present work.

In order to preserve local symmetries in a massive theory one could follow an approach already known in the spin-1 case. Instead of the usual Proca theory with Stueckelberg fields, we might follow \cite{cremmer} where the mass for the spin-1 field is generated via a gauge invariant coupling to a two-form field (anti-symmetric tensor). The Cremmer-Scherk or massive BF model (mBF henceforth) can be generalized to arbitrary dimensions and for the non Abelian case \cite{ft}. Inspired by duality relations, a spin-2 analogue of \cite{cremmer} has been suggested in \cite{ks}, henceforth we call it ``mBR'' model since the spin-1 curvature $F_{\mu\nu}(A)$ (field strength) is replaced by a linearized Riemann curvature $R_{\mu\nu\alpha\beta}^{(L)}(h)$ and the two-form field  $B_{\mu\nu}$ by a Riemann like tensor $B_{\mu\nu\alpha\beta}$.

 Here we investigate in detail the particle content of the mBR  model in terms of gauge invariants  and compare with the mBF spin-1 model, pointing out some important differences. We go beyond the linearized truncation of \cite{ks} and investigate possible nonlinear completions of the mBR  model in arbitrary $D$-dimensions. The linearized model in $D=3$ has inspired us to suggest a new class of bimetric models where however the particle sepectrum only contains massive spin-2 particles as we have checked at leading order in the decoupling limit. We show that the consistency of a possible nonlinear completion of the mBR  model in $D\ge 4$ is tightly connected with a consistent coupling to gravity of a higher rank massless spin-1 model suggested by Deser, Townsend and Siegel in \cite{dts}.

\section{Linearized mBR  model}

\subsection{From mBF to mBR}

In order to derive the linerarized massive mBR  model we start from the simpler spin-1 case by constructing a master action with two vector fields $(A_{\mu},f_{\mu})$, an antisymmetric tensor $B_{\mu\nu}=-B_{\nu\mu}$ and an arbitrary external source\footnote{We use $\eta_{\mu\nu}=(-,+,\cdots,+)$.} $J_{\mu}$,

\be S_{M}[J]=\int{d^{D}x}\Big\{-\frac{1}{4}F^{2}_{\mu\nu}(A)-\frac{m^{2}}{2}f_{\mu}f^{\mu}+\frac{m}{4}B^{\mu\nu}F_{\mu\nu}(f-A)+f_{\mu}J^{\mu}\Big\}\label{master1}.\ee

\no If we first integrate over $B_{\mu\nu}$ we arrive at the functional constraint $F_{\mu\nu}(f-A)=0$  whose general solution can be written in terms of a scalar Stueckelberg field $f_{\mu} = A_{\mu} + \p_{\mu}\phi/m$, now integrating over $f_{\mu}$ we obtain the Proca theory with a Stueckelberg field which is invariant under the $U(1)$ gauge transformations $(\delta A_{\mu}, \delta\phi) = (\p_{\mu}\Lambda,-m\, \Lambda)$,

\be S_{PS}[J]=\int{d^{D}x}\Big\{-\frac{1}{4}F^{2}_{\mu\nu}(A)-\frac{m^{2}}{2}\Big(A_{\mu}+\frac{\partial_{\mu}\phi}{m}\Big)^{2}+\Big(A_{\mu}+\frac{\partial_{\mu}\phi}{m}\Big)J^{\mu}\Big\}\label{ps}.\ee

\no If instead, we first integrate over $f_{\mu}$ we arrive at the massive BF (mBF) model in $D$ dimensions, up to quadratic terms in the external source,

\be S_{mBF}[J]=\int{d^{D}x}\Big\{-\frac{1}{4}F^{2}_{\mu\nu}(A)+\frac{1}{8}\partial_{\mu}B^{\mu\nu}\partial^{\lambda}B_{\lambda\nu}-\frac{m}{4}B^{\mu\nu}F_{\mu\nu}(A)  + B_{\mu}^*J^{\mu}+\mathcal{O}(J^{2})\Big\}\label{mbf},\ee

\no where $ B_{\mu}^*\equiv \frac{1}{m}\partial^{\nu}B_{\nu\mu}$. The mBF model is invariant under the independent scalar and transverse tensor ($\p^{\mu}\Lambda_{\mu\nu}^t =0$)  gauge transformations:

\bea \delta{A}_{\mu}=\partial_{\mu}\Lambda\qquad;\qquad\delta B_{\mu\nu}=\Lambda_{\mu\nu}^t \label{gt1}.\eea

\no In $D=3$ we can write $B_{\mu\nu} = \epsilon_{\mu\nu\rho}B^{\rho}$ and after rotating and decoupling the vector fields $B_{\mu}$ and $A_{\mu}$ it follows that the mBF model is equivalent to a couple of topologically massive  Maxwell-Chern-Simons (MCS) models \cite{djt} of opposite helicities $+1$ and $-1$ just like a Proca theory. In $D=4$ we can write $B^{\mu\nu}=\epsilon^{\mu\nu\rho\gamma}B_{\rho\gamma}$ and $mBF$ becomes the topologically massive $BF$ model also known as the Cremmer-Scherk model \cite{cremmer}.

The master action allows us to prove the duality between  $S_{PS}$ and $S_{mBF}$ via a  map between gauge invariant vectors that we read off from the linear terms in the sources, namely:  $A_{\mu} + \p_{\mu}\phi/m \leftrightarrow B_{\mu}^*$. The equations of motion $\delta S_{mBF}=0$ can be written as

\be \p^{\mu}F_{\mu\nu}[A] = m^2 B^*_{\nu} \quad ; \quad F_{\mu\nu}[A-B^*] = 0 \quad \label{mbfeom}. \ee

\no From the general solution of the second equation of (\ref{mbfeom}) $A_{\mu} = B_{\mu}^* + \p_{\mu}\Lambda$ back in the first equation  we have a Proca like equation $\p^{\mu}F_{\mu\nu}[B^*] = m^2 B^*_{\nu} $ which is equivalent to the Klein-Gordon equation $(\Box - m^2)B_{\mu}^* =0$ since $\p^{\mu}B^*_{\mu}=0$. Notice however, that the transverse condition is a dynamic equation in the Proca theory while it is a trivial identity for the dual field $B^*_{\mu}$.

 Now we can follow similar steps in order to derive the spin-2 version of $S_{mBF}$. First we define a master action in terms of two symmetric rank-2 tensors $(h_{\mu\nu},f_{\mu\nu})$ and a rank-4 tensor $B_{\mu\nu\alpha\beta}$ antisymmetric by the exchange $\mu \leftrightarrow \nu $ or $\alpha \leftrightarrow \beta $  and symmetric by $[\mu\nu]\leftrightarrow[\alpha\beta]$, like the Riemann tensor. By replacing the spin-1 curvature $F_{\mu\nu}$ by the linearized Riemann curvature tensor we have

\be S_{M}^{D}[T]=\int{d^{D}x}\Big\{-\frac{1}{2}h^{\mu\nu}G^{(L)}_{\mu\nu}(h)-\frac{m^{2}}{4}(f_{\mu\nu}f^{\mu\nu}-f^{2})+\frac{1}{2}B^{\mu\nu\alpha\beta}R^{(L)}_{\mu\nu\alpha\beta}(f-h) + f_{\mu\nu}T^{\mu\nu} \Big\}\label{master2},\ee

\no  where $R^{(L)}_{\alpha\mu\beta\nu}(h)=(\partial_{\mu}\partial_{\beta}h_{\alpha\nu}+\partial_{\alpha}\partial_{\nu}h_{\mu\beta}-\partial_{\mu}\partial_{\nu}h_{\alpha\beta}-\partial_{\alpha}\partial_{\beta}h_{\mu\nu})/2$.

 If we first integrate over $B_{\mu\nu\alpha\beta}$ we obtain the
 linearized zero  curvature condition

\be R^{(L)}_{\mu\nu\alpha\beta}(f-h)=0\label{zeroc},\ee

\no whose general solution gives rise to vector Stueckelberg fields,

\be f_{\mu\nu}(h,\psi)=h_{\mu\nu}+\frac{\partial_{\mu}\psi_{\nu}+\partial_{\nu}\psi_{\mu}}{m}\label{BRD12}.\ee

\no Back in the master action we have the Fierz-Pauli \cite{Fierz} theory with Stueckelberg fields which describes massive spin-2 particles
with linearized reparametrization invariance $(\delta h_{\mu\nu}, \delta\psi_{\mu})=(\p_{\mu}\epsilon_{\nu} + \p_{\nu}\epsilon_{\mu},-m\epsilon_{\mu})$, namely,

\be S_{FPS}^{D}[T]=\int{d^{D}x}\Big\{-\frac{1}{2}h^{\mu\nu}G^{(L)}_{\mu\nu}(h)-\frac{m^{2}}{4}\left\lbrack f_{\mu\nu}(h,\psi)f^{\mu\nu}(h,\psi)-f^{2}(h,\psi)\right\rbrack + f_{\mu\nu}(h,\psi)T^{\mu\nu} \Big\}\label{fps}\ee

\no On the other hand, by first integrating over  $f_{\mu\nu}$ in (\ref{master2}) one derives the massive BR (mBR) model \cite{ks}:

\be S^{D}_{mBR}[T]=\int{d^{D}x}\Big\{-\frac{1}{2}h^{\mu\nu}G^{L}_{\mu\nu}(h)-\frac{1}{2}B^{\mu\nu\alpha\beta}R^{(L)}_{\mu\nu\alpha\beta}(h)+\frac{1}{m^{2}}\mathcal{L}_{DTS}[B]+B^*_{\mu\nu}T^{\mu\nu}+\mathcal{O}(T^{2})\Big\}\label{mbr}\ee

\no where the dual symmetric tensor  is given by

\be
B^*_{\mu\nu}=-\,\frac{2}{m^{2}}\Big[\partial^{\alpha}\partial^{\beta}B_{\mu\alpha\nu\beta}-\frac{1}{(D-1)}\eta_{\mu\nu}\partial^{\alpha}\partial^{\beta}B_{\alpha\beta}\Big]\label{dualb}\ee


\no with $B^{\alpha\beta}\equiv\eta_{\mu\nu}B^{\mu\alpha\nu\beta}$ while the Lagrangian

\be \mathcal{L}^{D}_{DTS}[B]=(\partial_{\alpha}\partial_{\beta}B^{\mu\alpha\nu\beta})^{2}-\frac{1}{(D-1)}(\partial_{\alpha}\partial_{\beta}B^{\alpha\beta})^{2}\label{BRD16}\ee

\no  has been first obtained in \cite{dts} in $D=4$ and generalized to arbitrary dimensions in \cite{renato1}. It describes massless spin-1 particles like the Maxwell theory. Although of 4th order in derivatives, it is ghost free in arbitrary dimensions as shown in the appendix, see \cite{dts} for a proof in $D=4$. The DTS model is invariant under the following  scalar and tensor transformations:

\bea
\delta{B}_{\mu\alpha\nu\beta} = (\eta_{\mu\nu}\eta_{\alpha\beta}-\eta_{\mu\beta}\eta_{\nu\alpha})\pi +
\Lambda_{\mu\alpha\nu\beta} \label{gt2}
\eea
where $\Lambda_{\mu\alpha\nu\beta}$ must satisfy
\bea
\partial^{\mu}\partial^{\nu}\Lambda_{\mu\alpha\nu\beta}=0 \label{dtc}
\eea

\no The scalar transformation plays the role of the $U(1)$ symmetry of the Maxwell theory. Comparing with the spin-1 mBF model, we notice that the EH term (massless spin-2) is the analogue of the Maxwell theory (massless spin-1) while  ${\cal L}_{DTS}$ (massless spin-1) is the analogue of the antisymmetric model (massless spin-0). The spin-1 curvature $F_{\mu\nu}$ is replaced by the linearized Riemann tensor in the ``BR'' mixing term. Regarding the equations of motion $\delta S_{mBR} =0$, they can be written in close analogy with (\ref{mbfeom}),

\bea G_{\mu\nu}^{(L)}(h)& =& -\frac{m^2}2(B_{\mu\nu}^* - \eta_{\mu\nu} B^*) \quad , \label{mbreom1} \\
R_{\mu\nu\alpha\beta}^{(L)}(h-B^*) &=& 0 \quad \label{mbreom2} \eea

\no If we plug back in (\ref{mbreom1}) the general solution of (\ref{mbreom2}),
i.e., $h_{\mu\nu} = B_{\mu\nu}^* + (\p_{\mu}A_{\nu} + \p_{\nu}A_{\mu})/m$ we arrive exactly at Fierz-Pauli equations for the dual field :

\be   G_{\mu\nu}^L(B^*) = -\frac{m^2}2(B_{\mu\nu}^* - \eta_{\mu\nu} B^*) \quad , \label{fpeom} \ee

\no which leads to the Klein-Gordon equation $(\Box - m^2)B_{\mu\nu}^*=0$ and $B^*=0=\p^{\mu}B_{\mu\nu}^*$ which are typical equations for free massive spin-2 particles. Notice that $\p^{\mu}B_{\mu\nu}^*=\p_{\nu}B^*$
holds identically from the definition of the dual field, differently from the usual FP theory where it follows from the derivative of the FP equations for the fundamental field $h_{\mu\nu}$.

Since the equations of motion (\ref{fpeom}) are of fourth order in derivatives the skeptical reader may be questioning whether we have ghosts in the spectrum.
Also in the spin-1 case the Proca like equation in terms of the dual vector field $B_{\mu}^*$  is of 3rd order in derivatives. In order to make sure that in both $s=1$ and $s=2$ cases we only have the expected massive physical particles we perform in the next subsection a canonical analysis in terms of gauge invariant combinations of helicity variables in both mBF and mBR   models in arbitrary $D$ dimensions confirming our expectations.

\subsection{Canonical structure}

\subsubsection{The mBF model}

The canonical analysis of the  mBF model via Hamiltonian methods has appeared before in \cite{lahiri,kko} in $D=4$ and in \cite{wayne} in $D$ dimensions.
In \cite{lahiri,wayne} the Dirac method has been employed while \cite{kko} makes use of the Faddeev-Jackiw method. However, we believe that the purely Lagrangian  approach presented here is simpler and clarifies the role of each degree of freedom in producing the necessary mass terms. The Lorentz invariance is not explicit but we are able to write down the action only in terms of gauge invariants, there is no need of gauge fixing. Moreover, since Hamiltonian methods become cumbersome for higher derivative theories like the mBR   model, the spin-1 case works like an introduction and, more importantly, it points out that the analogy between the spin-1 and the spin-2 cases is not completely faithful.


We follow the approach used  for higher order higher spin theories in $D=3$ in \cite{da-hs}. The gauge invariants are built up in a constructive way from the definition of the  gauge transformations. For example, in the case of the $U(1)$ symmetry we take one of the $D$ equations $\delta A_{\mu} = \p_{\mu} \Lambda $ for the elimination of the  gauge parameter, explicitly we have\footnote{We assume vanishing fields at infinity, so $\nabla^2 A \equiv \p_j\p_j A =0  $ leads to $A =0$, where $A$  represents any field or its spacetime derivatives. Moreover, we use $T$ for space and $t$ for space-time transverse quantities.} : $\Lambda = \p_j \delta A_j/\nabla^2$. Plugging back in the $D-1$ remaining equations we obtain $D-1$ invariants $(\delta I_0(A),\delta I_j^T(A)) =(0,0) $ where $I_0(A)= A_0 - \p_0\p_j A_j/\nabla^2$ and  $I_j^T(A)=\theta_{jk}A_k$,
with the projection operator:

\be \theta_{jk}  \equiv \delta_{jk} -\frac{\p_j\p_k}{\nabla^2} \quad , \label{theta} \ee

\no The invariant $I_0(A)$ comes from the divergence of the electric field while  $I_j^T(A)$
is given in terms of the spatially transverse components of the magnetic field which are of course the only non vanishing ones. In general, the number of independent gauge invariants equals the number of independent fields minus the number of independent gauge parameters, symbolically,

\be N_I = N_{A} - N_{\Lambda} \label{ni} \quad . \ee

\no Thus, from the $D(D-1)/2$ equations $\delta{B}_{\mu\nu} = \Lambda_{\mu\nu}^t $ and the $D-1$ constraints $\partial^{\mu}\Lambda^t_{\mu\nu}=0$ we can
find $N_I = D(D-1)/2-[D(D-1)/2 - (D-1)]=(D-1)$
invariants\footnote{Specifically, we have decomposed a general antisymmetric field as
$(\Lambda_{0i},\Lambda_{ij} ) = (\lambda^{T}_{i}+\partial_{i}\varphi ,
 \Lambda^{T}_{ij}+\partial_{i}\omega^{T}_{j}-\partial_{j}\omega^{T}_{i})$ and used the transversality condition $\partial^{\mu}\Lambda^t_{\mu\nu}=0$ to get rid of redundant components
$(\varphi,\omega^{T}_{i})=(0,\dot\lambda^{T}_{i}/\nabla^{2})$}
 which turn out to be $ I_{\mu}^t(B) \equiv \p^{\nu}B_{\nu\mu}$.  Now we return to the mBF theory. Let us introduce an invertible decomposition without time derivatives in order to avoid changes in the canonical structure of the theory, namely,

\bea A_0 = \gamma \quad &;& A_j = v_j^T + \p_j \theta \label{adec} \\
B_{0j} = b_j^T + \p_j \psi \quad &;& \quad B_{ij} = b_{ij}^T +\p_i c_j^T - \p_j c_i^T \label{bdec} \eea

\no where $\p_j v_j^T = 0=\p_j b_j^T = \p_j c_j^T =\p_j b_{ij}^T=0$ and $b_{ij}^T=-b_{ji}^T$. So  we have

\bea {\cal L}_{mBF} &=&  \overbrace{\frac 12 \, v_j^T \, \Box \, v_j^T- \frac 12 \, (\gamma - \dot{\theta})\, \nabla^2 \, (\gamma - \dot{\theta})}^{-\frac 14 F_{\mu\nu}^2} \quad \overbrace{-\frac 12 \, \nabla^2 \psi \, \Box \, \psi + \frac 12 \, (\nabla^2c_j^T - \dot{b}_j^T)^2}^{+\frac 12 (\p^{\mu}B_{\mu\nu})^2}    \nn\\
&+& \underbrace{ m \, (\gamma - \dot{\theta})\, \nabla^2 \, \psi + m \, (\nabla^2c_j^T - \dot{b}_j^T) \, v_j^T }_{- \frac m2 \, B^{\mu\nu}F_{\mu\nu}}\label{mbffull}\eea

\no The free Lagrangian is a bilinear function of  the previously found  invariants:
$(I_0(A),I_j^T(A)) = (\gamma - \dot{\theta},v_j^T)$ and $(I_0(B),I_j^T(B))=(\nabla^2\psi,\nabla^2c_j^T - \dot{b}_j^T-\p_j\dot{\psi})$.
 In the Maxwell theory we have $D-2$ massless propagating modes ($v_j^T$) and 1 non propagating one ($\gamma - \dot{\theta}$), while in the antisymmetric tensor we have the opposite, 1 massless propagating mode ($\psi$) and $D-2$ non dynamic gauge invariants ($\nabla^2c_j^T - \dot{b}_j^T$). In the  BF term we have a perfect match. The non dynamic modes of one theory couple to the propagating ones of the other one in order to generate the mass terms. After diagonalizing the Lagrangian, the now massive $D-1$ propagating fields $(v_j^T,\psi)$ decouple  from the non propagating ones  giving rise to a massive spin-1 particle without extra fields or ghosts,

\be {\cal L}_{mBF} =  \frac 12 \, v_j^T \, (\Box - m^2) \, v_j^T + \frac 12 \, \psi \, (-\nabla^2) (\Box - m^2) \, \psi + \frac{C_j^T}2 - \frac 12 \, \Gamma \,\nabla^2 \,\Gamma \quad . \label{mbffinal} \ee

\no where  $(C_j^T,\Gamma)\equiv (\nabla^2c_j^T - \dot{b}_j^T+m\, v_j^T,\gamma-\dot{\theta}-m\, \psi)$ are the non dynamic invariants.

\subsubsection{The mBR   model}

Turning off the sources and redefining $B_{\mu\alpha\nu\beta}\to m B_{\mu\alpha\nu\beta}/2$, the Lagrangian for the linearized mBR   model
(\ref{mbr}) is given by:

\bea {\cal L}_{mBR} &=& \frac{1}{2}\left(\frac{1}{2}h_{\mu\nu}\Box h^{\mu\nu}-\frac{1}{2}h\Box
h+h\partial_{\mu}\partial_{\nu}h^{\mu\nu}
-h^{\mu\nu}\partial_{\mu}\partial^{\lambda}h_{\lambda\nu}\right) \nn\\
&+&\frac{1}{4}\big(\partial_{\mu}\partial_{\nu}B^{\mu\alpha\nu\beta}\,\big)^{2}-\frac{1}{4(D-1)}\big(\partial_{\alpha}\partial_{\beta}B^{\alpha\beta}\big)^{2} -\frac m4 B^{\mu\alpha\nu\beta}R^{(L)}_{\mu\alpha\nu\beta}(h) \label{lmbr}
\eea

 The linearized EH theory, first line of (\ref{lmbr}), is invariant under linearized reparametrizations $\delta{h}_{\mu\nu} = \partial_{\mu}\xi_{\nu}+\partial_{\nu}\xi_{\mu}$. We use $D$ of those $D(D+1)/2$ equations to determine the gauge parameters $\xi_{\mu}$ in terms of $\delta{h}_{\mu\nu} $ and plugging back the result in the transformations the remaining  equations furnish $N_I=D(D+1)/2-D=D(D-1)/2$ gauge invariants.
Using $(\xi_0,\xi_j) = (\omega,\xi_j^T+\p_j\xi)$ and  the decomposition:

\bea
h_{\mu\nu}\rightarrow\left\{\begin{array}{l}
h_{00} = \rho \\
h_{0i} = \gamma^{T}_{i}+\partial_{i}\theta \\
h_{ij} = {h}^{TT}_{ij}+\partial_{i}\psi^{T}_{j}+\partial_{j}\psi^{T}_{i}
+ \theta_{ij}\nabla^2 \psi + \p_i\p_j\phi
\end{array}\right.\quad,\quad
\eea
where the superscript \textit{TT} means transverse and traceless, i.e:
$\partial_{i}h^{TT}_{ij}=0=\delta_{ij}h^{TT}_{ij}$, we can write down the linearized EH theory  in terms of gauge invariants\footnote{Compare  with \cite{jmm} for $D=4$.}, up to total derivatives we have:

\be {\cal L}_{LEH} = \left(\sqrt{-g}R\right)_{hh} = \frac 14 h_{ij}^{TT} \, \Box \, h_{ij}^{TT} - \frac 12 \Gamma_j \, \nabla^2 \, \Gamma_j + \frac{(D-1)(D-2)}4\Phi \, \Box \, \Phi + \frac{(D-2)}2 \Phi \, R^{(L)} \label{lehdec} \ee

\no where the $D(D-3)/2$ invariants $h_{ij}^{TT}$ represent the propagating degrees of freedom of the graviton while the next $(D-2) + 2$ invariants:

\bea \Gamma_j^T &=& \gamma_j^T - \dot{\psi}_j^T \quad ; \quad \Phi = \nabla^2 \phi \label{gammaj} \\ R^{(L)} &=& \p^{\mu}\p^{\nu}h_{\mu\nu} -\Box \, h = \nabla^2 \, \left\lbrack \rho - 2\, \dot{\theta} + \ddot{\phi} + (2-D)\, \Box \psi \right\rbrack \label{R} \eea

\no are non propagating. In total we have the expected $D(D-1)/2$ gauge invariants. Notice that (\ref{gammaj}) and (\ref{R}) make clear that each invariant can be indeed associated with an independent field. In particular $\delta S_{LEH}/\delta\rho =0 $ establishes that the ``would be'' propagating mode $\Phi$ vanishes. Consequently, from $\delta S_{LEH}/\delta\Phi =0 $ we have vanishing scalar curvature ($R^{(L)}=0$) as expected from the Einstein equations in the vacuum.

Regarding the rest of the Lagrangian (\ref{lmbr}), we introduce the general decomposition:

\bea
B_{\mu\alpha\nu\beta}\rightarrow\left\{\begin{array}{l}
B_{[0i][0j]} = B^{T}_{ij}+\partial_{i}B^{T}_{j} + \partial_{j}B^{T}_{i}+2\partial_{i}\partial_{j} Z \\
B_{[0i][jk]} = B^{pT}_{i[jk]}+\partial_{j}C^{pT}_{ik}-\partial_{k}C^{pT}_{ij}
\\
B_{[ij][kl]} = B^{T}_{[ij][kl]}+ \left\lbrack\partial_{i}D^{T}_{j[kl]}+\partial_{k}D^{T}_{l[ij]}
+\partial_{i}(\partial_{k}W^{T}_{jl}-\partial_{l}W^{T}_{jk})+ (i\leftrightarrow{j},k\leftrightarrow{l})\right\rbrack \label{Bdec}\
\end{array}\right.
\eea

\no where $(B^{T}_{ij},W^{T}_{ij})=(B^{T}_{ji},W^{T}_{ji})$ while $C^{pT}_{ij}\ne C^{pT}_{ji}$. All spatial tensor  fields are transverse in all indices except the ones in the second line of (\ref{Bdec}) which are only partially transverse (pT), namely,  $\p_jB^{pT}_{i[jk]}=0=\partial_{j}C^{pT}_{ij}$ but $\p_iB^{pT}_{i[jk]}\ne 0$ and $\partial_{i}C^{pT}_{ij}\ne 0$. In the appendix we show that ${\cal L}_{DTS}$  in (\ref{BRD16})  can be written in terms of $(D+1)(D-2)/2$ gauge invariants, see (\ref{ldts}), which split into a transverse vector $V_j^T$ representing $D-2$ propagating massless modes, in agreement with a massless spin-1 field, plus $(D-1)(D-2)/2$ non propagating modes represented by a transverse symmetric spatial tensor $\mathbb{W}_{ij}^T$, which we further decompose into its trace $\mathbb{W}=\delta^{ij}\mathbb{W}^T_{ij}$ and its traceless and transverse piece $\omw_{ij} \equiv \mathbb{W}_{ij}^T - \theta_{ij}\mathbb{W}/(D-2)$, see  (\ref{Vj}) and (\ref{Wij}).

Regarding the $BR$ term, there is an important difference with the spin-1 case. Namely, the BF term does not break any of the symmetries (\ref{gt1}) of the rest of the model while the BR term breaks the scalar symmetry in (\ref{gt2}). Since we have one less gauge parameter we end up with one more gauge invariant besides $(V_j^T,\mathbb{W}_{ij}^T)$.
We find it by noticing that $\p^i\p^j \delta B_{[0i][0j]} = \p^i\p^j \Lambda_{[0i][0j]}=\p^{\mu}\p^{\nu} \Lambda_{[0\mu][0\nu]} =0 $ by virtue of
(\ref{dtc}). Back in the first line of (\ref{Bdec}) we have our last invariant $\delta Z =0$, which only appears in the BR term. Finally, from (\ref{lehdec}), (\ref{ldts}) and working out the BR term, we have for the whole mBR   model:

\bea {\cal L}_{mBR} &=& \frac 14 h_{ij}^{TT} \, \Box \, h_{ij}^{TT} - \frac 12 \Gamma_j^T \, \nabla^2 \, \Gamma_j^T + \frac{(D-1)(D-2)}4\Phi \, \Box \, \Phi + \frac{(D-2)}2 \Phi \, R^{(L)} \nn \\
&+& \frac 14 \left( \omw_{ij} \right)^2 +  \frac 12 V_j^T (-\nabla^2)\Box V_j^T + \frac {\mathbb{W}^2}{4\, (D-1)(D-2)} \nn\\
&+& \underbrace{\frac m2 \, \omw_{ij} \,  h_{ij}^{TT} + m\, \Gamma_j^T \, \nabla^2 V_j^T + \frac m2 \,  \mathbb{W} \, \Phi  + m\, Z \nabla^2 R^{(L)}}_{-\frac m4 B^{\mu\alpha\nu\beta}R^{L}_{\mu\alpha\nu\beta}(h)} \, .
\label{mbrfull}
\eea

Integrating over the non propagating fields $\omw_{ij},\Gamma_j^T$ and
$\mathbb{W}$ we generate mass terms for the helicities $\pm 2, \pm 1, 0$ of the massive spin-2 particle respectively,

\be {\cal L}_{mBR} = \frac 14 h_{ij}^{TT} \, (\Box -m^2)\, h_{ij}^{TT}  +  \frac 12 V_j^T (-\nabla^2)(\Box - m^2) V_j^T +  \frac{(D-1)(D-2)}4\Phi \, (\Box -m^2)\, \Phi \, , \label{mbrm}
\ee

\no while the integral over $Z$ produces the constraint
$R^{(L)}=0$ which eliminates $h_{00}$. We have not written down the non propagating modes in (\ref{mbrm}). They correspond to quadratic terms in field redefinitions of $\omw_{ij},\Gamma_j^T$ and $\mathbb{W}$. Thus, we have the same number of propagating and non propagating gauge invariant modes in the final Lagrangian just like in the spin-1 case. However,  the analogy between the mBF (\ref{mbffull}) and the mBR   model (\ref{mbrfull}) is not totally faithful. Although one might think at first sight of a perfect match between propagating and non propagating  modes $( h_{ij}^{TT},V_j^T,\Phi) \leftrightarrow (\omw_{ij},\Gamma_j^T,\mathbb{W})$, we recall that $\Phi$ is a non propagating gauge invariant in the LEH theory, it only becomes a dynamic field because of
the last term in (\ref{mbrfull}) which annihilates the effect of the term $\Phi \, R^{(L)} $ which sets $\Phi =0$. So in the spin-2 case the role of the BR term is not only to couple dynamic with non dynamic fields but also to turn a non propagating mode into a propagating one and this is only possible because the scalar symmetry ( $U(1)$ symmetry) of the DTS model is broken giving rise to the $Z$ invariant.

Since the linearized mBR model is a consistent ghost free description of massive spin-2 particles with linearized reparametrization invariance, it is natural to try to go beyond the free theory and develop a consistent non linear version of the model invariant under general coordinate transformation. In order to prepare the ground for an analysis via decoupling limit, see \cite{Hinter-1,deRham-1}, in the next subsection we take a closer look at the pure massless limit starting with  the simpler $D=3$ case.

\subsection{The $D=3$ case and the massless limit}

 Since the mBR model describes massive spin-2 particles with reparemetrization invariance, it might be related with the New Massive Gravity (NMG) theory of \cite{bht} for $D=3$. Indeed, see \cite{ks}, in $D=3$ we can write the Riemann tensor in terms of the Einstein tensor $ R_{\mu\nu\alpha\beta}=\epsilon_{\mu\nu\lambda}\epsilon_{\alpha\beta\sigma}G^{\lambda\sigma}$. So one  can write down the  mBR  model (\ref{master2}) in $D=3$, without sources, in terms of three symmetric rank-2 tensors:

\bea S^{D=3}_{mBR}&=&\int{d^{3}x}\Big\{-\frac{1}{2}h^{\mu\nu}G^{(L)}_{\mu\nu}(h)-\frac{m^{2}}{4}(f^{2}_{\mu\nu}-f^{2}) + \tb^{\mu\nu}G^{(L)}_{\mu\nu}(f-h)\Big\}\label{BRD=35a}\\ &=&\int{d^{3}x}\Big\{-\frac{1}{2}h_+^{\mu\nu}G^{(L)}_{\mu\nu}(h_+)+\frac{1}{2}\tb^{\mu\nu}G^{(L)}_{\mu\nu}(\tb)-\frac{m^{2}}{4}(f^{2}_{\mu\nu}-f^{2})+f^{\mu\nu}G^{(L)}_{\mu\nu}(\tb)\Big\},\label{BRD=35b}\eea
where  $(h_+)_{\mu\nu} \equiv h_{\mu\nu} + \tb_{\mu\nu} $ and
 we have introduced the following symmetric tensor which has the same number of independent components of $B_{\alpha\beta\lambda\sigma}$ in $D=3$ :
\be \tb^{\mu\nu}\equiv\frac{1}{2}\epsilon^{\mu\alpha\beta}\epsilon^{\nu\lambda\sigma}B_{\alpha\beta\lambda\sigma} = \tb^{\nu\mu}\label{BRD=36}\ee

%
%
\no Integrating over $f_{\mu\nu}$ in (\ref{BRD=35b}) and neglecting the first EH term which decouples and has no particle content in $D=3$, we obtain the linearized version of the NMG theory \cite{bht}:
\be S^{D=3}_{mBR}=\int{d^{3}x}\Big\{+\frac{1}{2}\tb^{\mu\nu}G^{(L)}_{\mu\nu}(\tb)+\frac{1}{m^{2}}\mathcal{L}_{K}
[\tb]\Big\}\label{lnmg}\ee
with

\bea \mathcal{L}_{K}&=&
\frac{1}{4}\square\tb_{\mu\nu}\square\tb^{\mu\nu}-\frac{1}{2}\square\tb_{\mu\nu}\partial^{\mu}\partial_{\alpha}\tb^{\alpha\nu}+\frac{1}{8}(\partial_{\mu}\partial_{\nu}\tb^{\mu\nu})^{2}+\frac{1}{4}\square\tb\partial_{\mu}\partial_{\nu}\tb^{\mu\nu}-\frac{1}{8}(\square\tb)^{2}
\label{BRD=313}\\ &=&\Big[R^{\,2}_{\mu\nu}(\gamma)-\frac{3}{8}R(\gamma)^{\,2}\Big]_{\tb\tb}
\label{lkrr}\eea

\no We have considered $\tb_{\mu\nu}$ as the fluctuation of a metric about the flat space: $\gamma_{\mu\nu}=\eta_{\mu\nu}-\tb_{\mu\nu}$. The first term in (\ref{lnmg}) becomes the linearized version of the EH theory
with the sign opposite to the usual one. As a first step to check the particle content of the NMG model beyond the linearized theory ({\ref{lnmg}) one might try to look at the leading order in the decoupling limit where $m\to 0$ and $m_p \to \infty$  as we later explain, see \cite{yavin}. Notice however, that
the fourth order theory (\ref{lnmg}) is singular at $m\to 0$. This is also the case of  the  D-dimensional mBR  model in (\ref{mbr}). If instead, we simply abandon the canonical mass dimension and redefine  $B_{\mu\alpha\nu\beta} \to m\, B_{\mu\alpha\nu\beta}$, when $m\to 0$ we will be left only with ${\cal L}_k$ which is equivalent to the Maxwell theory in $D=3$ and has $D-2=1$ degree of freedom instead of two degrees of freedom as expected for a parity doublet of spin-2 in $D=3$. If we try $m\to 0$ directly in the 2nd order theories (\ref{BRD=35a}) or (\ref{BRD=35b}) we have zero degrees of freedom. So withouth extra fields we do not have a smooth massless limit. This point has been investigated in \cite{yavin}. Going back to the second order version\footnote{We have neglected the first term of (\ref{BRD=35b}) which has no particle content.}  (\ref{BRD=35b}),

\be S_{mBR}^{(D=3)}=\int{d^{3}x}\Big\{+\frac{1}{2}\tb^{\mu\nu}G^{(L)}_{\mu\nu}(\tb)-f^{\mu\nu}G^{(L)}_{\mu\nu}(\tb)-\frac{m^{2}}{4}(f^{2}_{\mu\nu}-f^{2})\Big\}\label{BHT12} \quad , \ee

\no we notice that the mass term breaks the symmetry
 $(\delta{f}_{\mu\nu},\delta{h}_{\mu\nu})=(\partial_{\mu}\zeta_{\nu}+\partial_{\nu}\zeta_{\mu},0)$. In order to recover it one introduces auxiliary  compensating (Stueckelberg) fields via:
\be f_{\mu\nu}\;\rightarrow\; f_{\mu\nu}+\frac{\partial_{\mu}A_{\nu}+\partial_{\nu}A_{\mu}}{m}+\frac{2\,\partial_{\mu}\partial_{\nu}\phi}{m^{2}}\label{stuck1}\ee
%

\no After taking $m\rightarrow0$ one has:
\be S_{mBR}^{(D=3)}(m\rightarrow0)=\int{d^{3}x}\Big\{+\frac{1}{2}\tb^{\mu\nu}G^{(L)}_{\mu\nu}(\tb)-f^{\mu\nu}G^{(L)}_{\mu\nu}(\tb)-\frac{1}{4}F^{\,2}_{\mu\nu}(A)-f_{\mu\nu}(\partial^{\mu}\partial^{\nu}\phi-\eta^{\mu\nu}\square\phi)\Big\}\label{BHTlmn7}\ee

\no Integrating over  $f_{\mu\nu}$ one obtains:
\be G^{(L)}_{\mu\nu}(\tb)=\eta_{\mu\nu}\square\phi-\partial_{\mu}\partial_{\nu}\phi\label{BHTlmn8}\ee

\no whose general solution is given by
\be \tb_{\mu\nu}=2\eta_{\mu\nu}\phi+\mbox{pure gauge}\label{BHTlmn9}\ee

\no Back in (\ref{BHTlmn7}) one finds \cite{yavin}:
\be S_{mBR}^{(D=3)}(m\rightarrow0) = \int{d^{3}x}\Big\{-\frac{1}{4}F^{\,2}_{\mu\nu}(A)+2\phi\square\phi\Big\}\label{BHTlmn10}\ee

\no The Maxwell theory in $D=3$ has only one degree of freedom, it is dual to a scalar field,  we end up with a total of two degrees of freedom which is the expected number of degrees of a parity invariant massive spin-2 field  in $D=3$ (parity doublet). So we have a smooth massless limit. This is similar to the spin-1 master action (\ref{master1}) after $f_{\mu} \to f_{\mu} + \p_{\mu}\phi/m $ and $m\to 0$ which leads to Maxwell plus a scalar field again, corresponding to $D-2+1=D-1$ degrees of freedom, in agreement with the Proca theory.

Now we go back to the mBR  model in $D$-dimensions in its second order version (\ref{master2}). After inserting Stueckelberg fields  we have

\bea S_{BRStuec}^{D}&=&\int{d^{D}x}\Big\{-\frac{1}{2}h^{\mu\nu}G^{(L)}_{\mu\nu}(h)-\frac{1}{2}B^{\mu\nu\alpha\beta}R^{(L)}_{\mu\nu\alpha\beta}(h)+\frac{1}{2}B^{\mu\nu\alpha\beta}R^{(L)}_{\mu\nu\alpha\beta}(f)-\frac{1}{4}F^{2}_{\mu\nu}(A)\nn\\
&&\qquad\quad-\frac{m^{2}}{4}(f_{\mu\nu}f^{\mu\nu}-f^{2})-mf_{\mu\nu}(\partial^{\mu}A^{\nu}-\eta^{\mu\nu}\partial_{\alpha}A^{\alpha})-f_{\mu\nu}(\partial^{\mu}\partial^{\nu}\phi-\eta^{\mu\nu}\square\phi)\Big\}\nn\\
\label{LmnBRD4}\eea

\no After $m\rightarrow0$ we obtain up to total derivatives,
\bea S^{D}_{BR}(\,m\rightarrow0)&=&\int{d^{D}x}\Big\{-\frac{1}{2}h^{\mu\nu}G^{(L)}_{\mu\nu}(h)+h^{\mu\nu}\partial^{\alpha}\partial^{\beta}B_{\mu\alpha\nu\beta}-f^{\mu\nu}\partial^{\alpha}\partial^{\beta}B_{\mu\alpha\nu\beta}-\frac{1}{4}F^{2}_{\mu\nu}(A)\nn\\
&&\qquad\qquad-f_{\mu\nu}(\partial^{\mu}\partial^{\nu}\phi-\eta^{\mu\nu}\square\phi)\Big\}\label{LmnBRD8}\eea

\no Integrating over the field $B_{\mu\alpha\nu\beta}$ we have

\be R^{(L)}_{\mu\nu\alpha\beta}(f) = R^{(L)}_{\mu\nu\alpha\beta}(h) \label{rfrh} \ee

\no whose general solution is given by $f_{\mu\nu} = h_{\mu\nu} + \p_{\mu}\chi_{\nu} + \p_{\nu}\chi_{\mu} $, where  $\chi_{\mu}$ are arbitrary pure gauge vector fields. Back  in (\ref{LmnBRD8}) we get :
\bea S^{D}_{\,m\rightarrow0}&=&\int{d^{D}x}\Big\{-\frac{1}{2}h^{\mu\nu}G^{(L)}_{\mu\nu}(h)+h^{\mu\nu}(\eta_{\mu\nu}\square\phi-\partial_{\mu}\partial_{\nu}\phi)-\frac{1}{4}F^{2}_{\mu\nu}(A)\Big\}\label{LmnBRD14}\eea

\no After a conformal redefinition  $h_{\mu\nu}\,\rightarrow\,\bar{h}_{\mu\nu}+\frac{2}{(D-2)}\eta_{\mu\nu}\phi$ we have a diagonal form which splits into spin-2, spin-1 and spin-0 sectors,
\bea S^{D}_{\,m\rightarrow0}&=&\int{d^{D}x}\Big\{-\frac{1}{2}\bar{h}^{\mu\nu}G^{(L)}_{\mu\nu}(\bar{h})-\frac{1}{4}F^{2}_{\mu\nu}(A)+\frac{(D-1)}{(D-2)}\phi\square\phi\Big\}\label{LmnBRD15}\eea

In $D$ dimensions the linearized  Einstein-Hilbert term possess  $D(D-3)/2$ degrees of freedom which altogether with  $(D-2) +1$ degrees of freedom of the vector and scalar sectors lead to  $(D+1)(D-2)/2$ which is the same number of independent modes of the massive spin-2 Fierz-Pauli \cite{Fierz} theory described by a symmetric rank-2 tensor ($D(D+1)/2$) constrained by the  $(D+1)$ FP conditions:
$\p^{\mu}h_{\mu\nu} = 0 = \eta^{\mu\nu} h_{\mu\nu}$. At $D=4$ we end up with $5=2s+1$ degrees of freedom as expected. In summary, the $D$-dimensional mBR  model with Stueckelberhg fields via (\ref{stuck1}) has a smooth massless limit like the FP theory with Stueckelberg fields.

\section{Nonlinear mBR  models}

Before we search for non linear extensions of the $D$-dimensional mBR  model we have found convenient again to first address the $D=3$ case in the next subsection which is simpler and has allowed us to suggest a new class of bimetric models.

\subsection{A new class of $D=3$ bimetric models}

A closer look at the linearized $D=3$ mBR  model (\ref{BRD=35a})

\be S^{D=3}_{BR}(\tb,h,f) =\int{d^{3}x}\Big\{-\frac{1}{2}h^{\mu\nu}G^{(L)}_{\mu\nu}(h)-\tb^{\mu\nu}G^{(L)}_{\mu\nu}(h)-\frac{m^{2}}{4}(f^{2}_{\mu\nu}-f^{2})+f^{\mu\nu}G^{(L)}_{\mu\nu}(B)\Big\}\quad , \label{BRD=35c} \ee,

\no inspires us to introduce two metrics $(g_{\mu\nu},\gamma_{\mu\nu}) = (\eta_{\mu\nu} + h_{\mu\nu},\eta_{\mu\nu} + B_{\mu\nu})$. While the last two terms give rise, after integrating over $f_{\mu\nu}$, to ${\cal L}_K$ which on its turn can be nonlinearly completed in terms of squares of curvatures,  the first and second terms can be written as linearizations of linear combinations of $\sqrt{-g}R(g) $ and $\sqrt{-g}\gamma^{\mu\nu}G_{\mu\nu}(g)$ where
$\gamma^{\mu\nu}=g^{\mu\alpha}g^{\nu\beta}\gamma_{\alpha\beta}$. The field $f_{\mu\nu}$  remains an auxiliary field. In fact we can go beyond (\ref{BRD=35c}) and suggest a quite general Ansatz for a new class of bimetric models in $D=3$, namely, restoring the Planck mass in the action we have

\bea S_{g-\gamma}&=&M_{P}\int{d^{3}x}\Big\{a\sqrt{-g}R(g)+b\sqrt{-\gamma}R(\gamma)+c\sqrt{-g\,}\gamma^{\mu\nu}G_{\mu\nu}(g)+d\sqrt{-\gamma\,}g^{\mu\nu}G_{\mu\nu}(\gamma)\nn\\
&&\qquad\qquad\;+k\sqrt{-\gamma}\widehat{f}^{\mu\nu}G_{\mu\nu}(\gamma)-\frac{m^{2}}{4}\sqrt{-\gamma}(\widehat{f}^{\,2}_{\mu\nu}-\widehat{f}^{\,2})\Big\}\label{bBR1}\eea

\no where $a$, $b$, $c$, $d$ and $k$ are arbitrary constants and $\widehat{f}_{\mu\nu}$ are auxiliary fields such that $\widehat{f}^{\mu\nu}=\gamma^{\mu\alpha}\gamma^{\nu\beta}\widehat{f}_{\alpha\beta}$ and $\widehat{f}=\gamma^{\mu\nu}\widehat{f}_{\mu\nu}$. The simultaneous flat space solution: $g_{\mu\nu}=\gamma_{\mu\nu}=\eta_{\mu\nu}$ ; $\widehat{f}_{\mu\nu}=0$ solves in general  the equations of motion of (\ref{bBR1}). Introducing

\be g_{\mu\nu}=\eta_{\mu\nu}+\frac{h_{\mu\nu}}{\sqrt{M_{P}}}\quad,\quad \gamma_{\mu\nu}=\eta_{\mu\nu}+\frac{b_{\mu\nu}}{\sqrt{M_{P}}}\quad,\quad \widehat{f}_{\mu\nu}=\frac{f_{\mu\nu}}{\sqrt{M_{P}}}\label{bBR8}\ee

\no At quadratic order in  $D=3$ we have:

\bea M_P\sqrt{-g}R(g)\;&\rightarrow&\;-\frac{1}{2}h^{\mu\nu}G^{(L)}_{\mu\nu}(h)\label{Expan1}\\
M_P\sqrt{-\gamma}R(\gamma)\;&\rightarrow&\;-\frac{1}{2}b^{\mu\nu}G^{(L)}_{\mu\nu}(b)\label{Expan2}\\
M_P\sqrt{-g}\gamma^{\mu\nu}G_{\mu\nu}(g)\;&\rightarrow&\;-b^{\mu\nu}G^{(L)}_{\mu\nu}(h)+\frac{5}{4}h^{\mu\nu}G^{(L)}_{\mu\nu}(h)\label{Expan3}\\
M_P\sqrt{-\gamma}g^{\mu\nu}G_{\mu\nu}(\gamma)\;&\rightarrow&\;-h^{\mu\nu}G^{(L)}_{\mu\nu}(b)+\frac{5}{4}b^{\mu\nu}G^{(L)}_{\mu\nu}(b)\label{Expan4}\\
M_P\sqrt{-\gamma}\widehat{f}^{\mu\nu}G_{\mu\nu}(\gamma)\;&\rightarrow&\; f^{\mu\nu}G^{(L)}_{\mu\nu}(b)\label{Expan5}\\
M_P\sqrt{-\gamma}(\widehat{f}^{\,2}_{\mu\nu}-\widehat{f}^{\,2})\;&\rightarrow&\;(f^{\,2}_{\mu\nu}-f^{\,2})\label{Expan6}\eea
On the right side all indices are raised with the Minkowisky metric $\eta^{\mu\nu}$. Using (\ref{Expan1}-\ref{Expan6}) in (\ref{bBR1}), we obtain at quadratic order:
\bea S&=&\int{d^{3}x}\Big\{-\frac{r}{2}h^{\mu\nu}G^{(L)}_{\mu\nu}(h)-\frac{s}{2}b^{\mu\nu}G^{(L)}_{\mu\nu}(b)-tb^{\mu\nu}G^{(L)}_{\mu\nu}(h)+kf^{\mu\nu}G^{(L)}_{\mu\nu}(b)\nn\\
&&\qquad\qquad-\,\frac{m^{2}}{4}(f^{\,2}_{\mu\nu}-f^{\,2})\Big\}\label{bBR12}\eea
where:
\be r=a-\frac{5}{2}c \quad,\quad s=b-\frac{5}{2}d \quad,\quad t=c+d\label{bBR14}\ee

\no Since the EH theory has no content in $D=3$ we  must have $k\ne 0$ in order to avoid the empty content of the first three terms of (\ref{bBR12}). If  $s=0=t$  the first term decouples from the $b_{\mu\nu}$ field and becomes the quadratic truncation of the EH term and the last two terms are equivalent to the ``K'' term of (\ref{lkrr})  which is equivalent to the Maxwell theory as shown in \cite{more} with only one degree of freedom in $D=3$. So henceforth we assume that $k\ne 0$ and $s$ or $t$ must be non vanishing.  If $r=0$ we must have $t=0$ otherwise the $h_{\mu\nu}$ equation of motion leads to $b_{\mu\nu}$ pure gauge and we would have no content again. Moreover if $r=0=t$ we must have $s<0$ in order that the EH term for the b-field has the ``wrong sign'', typical of the NMG model \cite{bht}, which guarantees a physical massive spin-2 particle. If $r\ne 0$ we can diagonalize the first three terms and write

\bea S&=&\int{d^{3}x}\Big\{-\frac{r}{2}\mathcal{H}^{\mu\nu}G^{(L)}_{\mu\nu}(\mathcal{H})+\frac{1}{2}\Big(\frac{t^{2}}{r}-s\Big)b^{\mu\nu}G^{(L)}_{\mu\nu}(b)+k\,f^{\mu\nu}G^{(L)}_{\mu\nu}(b)\nn\\
&&\qquad\qquad-\frac{m^{2}}{4}(f^{\,2}_{\mu\nu}-f^{\,2})\Big\}\label{bBR16}\eea
Where $\mathcal{H}_{\mu\nu}\equiv h_{\mu\nu}+\frac{t}{r}b_{\mu\nu}$. Since the first term (linearized Einstein-Hilbert) has no particle content, we may keep only the remaining ones which have the form of a second order formulation of the NMG model, namely,

\bea S&=&\int{d^{3}x}\Big\{+\frac{1}{2}\Big(\frac{t^{2}}{r}-s\Big)b^{\mu\nu}G^{(L)}_{\mu\nu}(b)+k\,f^{\mu\nu}G^{(L)}_{\mu\nu}(b)-\frac{m^{2}}{4}(f^{\,2}_{\mu\nu}-f^{\,2})\Big\}\label{bBR19}\eea

\no The particle content of (\ref{bBR19}) corresponds to a physical massive spin-2 particle with two helicity states $\pm 2$  whenever the EH coefficient has the ``wrong sign'' and $k\ne 0$,

\be \Big(\frac{t^{2}}{r}-s\Big)>0 \qquad ; \qquad  k\ne 0 \label{bBR21}\ee

\no Comparing with (\ref{bBR1}) we identify the linearized mBR  model (\ref{BRD=35c}) corresponds to  $a=7/2;c=1;b=d=0$ and $k=1$ which implies $(r,s,t)=(1,0,1)$.

%

Now in order to go beyond the linearized approximation we investigate now the decoupling limit of (\ref{bBR1}) at leading order. Namely, following \cite{yavin} we will take the following double limit while keeping
the scale $\Lambda_{5/2}$  fixed,

\be m\rightarrow0\quad,\quad M_{P}\rightarrow\infty\quad,\quad \Lambda_{5/2}\equiv (\sqrt{M_{P}}m^{2})^{2/5} \label{LDbBR7}\ee

\no Regarding the notation, we use $\widetilde\nabla_{\mu}$   for the covariant derivative with respect to the metric $\gamma_{\mu\nu}$ while $\nabla_{\mu}$ corresponds to the metric $g_{\mu\nu}$. Similarly to \cite{yavin}, we first notice that the term $\sqrt{-\gamma}\widehat{f}^{\mu\nu}G_{\mu\nu}(\gamma)$ is invariant under  $\delta\widehat{f}_{\mu\nu}=\widetilde{\nabla}_{\mu}\zeta_{\nu}+\widetilde{\nabla}_{\nu}\zeta_{\mu}$ which is broken by the mass term for the auxiliary fields $\widehat{f}_{\mu\nu}$.  In order to restore the symmetry we substitute in (\ref{bBR1}):
\be \widehat{f}_{\mu\nu}\;\rightarrow\; \frac{f_{\mu\nu}}{\sqrt{M_{P}}}+\frac{\widetilde\nabla_{\mu}A_{\nu}+\widetilde\nabla_{\nu}A_{\mu}}{\sqrt{M_{P}}m}+2\frac{\widetilde\nabla_{\mu}\widetilde\nabla_{\nu}\phi}{\sqrt{M_{P}}m^{2}}\label{LDbBR1}\ee
Up to total derivatives we obtain:
\bea
S_{g-\gamma}&=&\int{d^{3}x}\Big\{a\,M_{P}\sqrt{-g}R(g)+b\,M_{P}\sqrt{-\gamma}R(\gamma)+c\,M_{P}\sqrt{-g}\gamma^{\mu\nu}G_{\mu\nu}(g)\nn\\
&&\qquad\quad+d\,M_{P}\sqrt{-\gamma}g^{\mu\nu}G_{\mu\nu}(\gamma)+k\sqrt{M_{P}}\sqrt{-\gamma}f^{\mu\nu}G_{\mu\nu}(\gamma)-\frac{m^{2}}{4}\sqrt{-\gamma}(\widehat{f}^{\,2}_{\mu\nu}-\widehat{f}^{\,2})\nn\\
&&\qquad\quad-m\sqrt{-\gamma}f^{\mu\nu}(\widetilde\nabla_{\mu}A_{\nu}-\gamma_{\mu\nu}\widetilde\nabla_{\alpha}A^{\alpha})-\sqrt{-\gamma}f^{\mu\nu}(\widetilde\nabla_{\mu}\widetilde\nabla_{\nu}\phi-\gamma_{\mu\nu}\widetilde\nabla^{2}\phi)\nn\\
&&\quad\qquad-\,\frac{1}{2}\sqrt{-\gamma}\Big(\widetilde\nabla_{\mu}A_{\nu}\widetilde\nabla^{\mu}A^{\nu}-\widetilde\nabla_{\mu}A^{\mu}\widetilde\nabla_{\nu}A^{\nu}\Big)+\frac{1}{2}\sqrt{-\gamma}R_{\mu\nu}(\gamma)A^{\mu}A^{\nu}\nn\\
&&\qquad\quad+2\frac{\sqrt{-\gamma}}{m}R_{\mu\nu}(\gamma)A^{\mu}\widetilde\nabla^{\nu}\phi+\frac{\sqrt{-\gamma}}{m^{2}}R_{\mu\nu}(\gamma)\widetilde\nabla^{\mu}\phi\widetilde\nabla^{\nu}\phi\Big\}\label{LDbBR3}\eea
where we have used
$ [\widetilde\nabla_{\mu},\widetilde\nabla_{\nu}]V^{\mu}=R_{\mu\nu}(\gamma)V^{\mu}$ with $V^{\mu} = A^{\mu}$ and $\widetilde\nabla^{\mu}\phi $.


Using (\ref{bBR8}) and taking the decoupling limit we have:

\bea S\,'_{g-\gamma}&=&\int{d^{3}x}\Big\{-\frac{r}{2}h^{\mu\nu}G^{(L)}_{\mu\nu}(h)-\frac{s}{2}b^{\mu\nu}G^{(L)}_{\mu\nu}(b)-tb^{\mu\nu}G^{(L)}_{\mu\nu}(h)-\frac{1}{4}F^{\,2}_{\mu\nu}(A)\nn\\
&&\qquad\quad+kf^{\mu\nu}G^{(L)}_{\mu\nu}(b)-f^{\mu\nu}(\partial_{\mu}\partial_{\nu}\phi-\eta_{\mu\nu}\square\phi)+\frac{1}{\Lambda^{5/2}_{5/2}}R^{(L)}_{\mu\nu}(b)\partial^{\mu}\phi\partial^{\nu}\phi\Big\}\nn\\
\label{LDbBR10}\eea

\no Integrating over $f_{\mu\nu}$ which appears linearly we obtain
\be k\,G^{(L)}_{\mu\nu}(b)=\partial_{\mu}\partial_{\nu}\phi-\eta_{\mu\nu}\square\phi\label{LDbBR12}\ee
whose general solution is:
\be b_{\mu\nu}=-\,\frac{2}{k}\eta_{\mu\nu}\phi\,+\partial_{\mu}\zeta_{\nu}+\partial_{\nu}\zeta_{\mu}\label{LDbBR14}\ee

\no Back in (\ref{LDbBR10}),
\bea S\,'_{g-\gamma}&=&\int{d^{3}x}\Big\{-\frac{r}{2}h^{\mu\nu}G^{(L)}_{\mu\nu}(h)-\frac{2s}{k^{2}}\phi\square\phi-\frac{t}{k}h^{\mu\nu}(\partial_{\mu}\partial_{\nu}\phi-\eta_{\mu\nu}\square\phi)-\frac{1}{4}F^{\,2}_{\mu\nu}(A)\nn\\
&&\qquad\quad+\frac{1}{k\Lambda^{5/2}_{5/2}}\Big(\square\phi\partial_{\mu}\phi\partial^{\mu}\phi+\partial_{\mu}\partial_{\nu}\phi\partial^{\mu}\phi\partial^{\nu}\phi\Big)\Big\}\label{LDbBR17}\eea

\no After a conformal field redefinition:
\be h_{\mu\nu}\;\rightarrow\;h_{\mu\nu}+\frac{2t}{kr}\eta_{\mu\nu}\phi\label{LDbBR19}\ee
We have a diagonal action:

\bea S\,'_{g-\gamma}=\int{d^{3}x}\Big\{-\frac{r}{2}h^{\mu\nu}G^{(L)}_{\mu\nu}(h)-\frac{1}{4}F^{\,2}_{\mu\nu}(A) +\frac{2}{k^{2}}\Big(\frac{t^{2}}{r}-s\Big)\phi\square\phi
+\frac{1}{2k\Lambda^{5/2}_{5/2}}\square\phi\partial_{\mu}\phi\partial^{\mu}\phi\Big\}\label{LDbBR20}\eea

\no So, as in \cite{yavin}, we have a smooth massless limit where the scalar field has Galileon self-interaction and quadratic kinetic term with the correct sign in agreement with (\ref{bBR21}). The EH term has no content while the Maxwell theory has one degree of freedom.  So the decoupling limit of the bimetric model  at leading order is free of ghosts as far as (\ref{bBR21}) holds true.
Clearly, a complete nonlinear analysis including a detailed study of the Hamiltonian structure of the model (\ref{bBR1}), as the one carried out in \cite{hr} for the Hassan and Rosen 4D bimetric model, is required for a full consistency proof.

In a more general setting we can add cosmological terms to (\ref{bBR1}) like $\Lambda_g\sqrt{-g} + \Lambda_{\gamma}\sqrt{-\gamma}$ and look for AdS solutions. It is possible to show that there is always a region in the parameters space of the model for which AdS solutions with proportional metrics $g_{\mu\nu} \propto \gamma_{\mu\nu} $ do exist.

\subsection{Searching for a single metric mBR  model}

Now we come back to the linearized $D=3$ mBR  model in (\ref{BRD=35c}) and consider  the $\tb_{\mu\nu}$ field as an extra field appearing linearly in the action instead of a second metric fluctuation in order to prepare the ground for $D>3$. The only metric now is $g_{\mu\nu}=\eta_{\mu\nu} + h_{\mu\nu}$. We arrive at the natural non linear generalization:

\bea S^{D=3}_{NLBR}&=&M_{P}\int{d^{3}x}\sqrt{-g}\Big\{R(g)-\widetilde{B}^{\mu\nu}G_{\mu\nu}(g)+\widetilde{B}^{\mu\nu}\mathbb{G}_{\mu\nu}(\widehat{f})-\frac{m^{2}}{4}(\widehat{f}^{\,2}_{\mu\nu}-\widehat{f}^{2})\Big\}\nn\\
\label{VnlBRD=33}\eea
where
\bea \mathbb{G}_{\mu\nu}(\widehat{f})&\equiv&\frac{1}{2}\Big(-\nabla^{2}\widehat{f}_{\mu\nu}+\frac{1}{2}\nabla_{\mu}\nabla^{\alpha}\widehat{f}_{\alpha\nu}+\frac{1}{2}\nabla^{\alpha}\nabla_{\mu}\widehat{f}_{\alpha\nu}+\frac{1}{2}\nabla_{\nu}\nabla^{\alpha}\widehat{f}_{\alpha\mu}+\frac{1}{2}\nabla^{\alpha}\nabla_{\nu}\widehat{f}_{\alpha\mu}\nn\\
&&\quad\;-\nabla_{\mu}\nabla_{\nu}\widehat{f}+g_{\mu\nu}\nabla^{2}\widehat{f}-g_{\mu\nu}\nabla_{\alpha}\nabla_{\beta}\widehat{f}^{\alpha\beta}\Big)\label{VnlBRD=35}\eea
The symbol $\nabla_{\mu}$ stands for the usual covariant derivative with respect to $g_{\mu\nu}$ and we use the notation $\nabla^{2}=\nabla_{\mu}\nabla^{\mu}$. The Einstein-like tensor $\mathbb{G}_{\mu\nu}$ satisfies:
\be \widetilde{B}^{\mu\nu}\mathbb{G}_{\mu\nu}(\widehat{f})=\widehat{f}^{\mu\nu}\mathbb{G}_{\mu\nu}(\widetilde{B})+\mbox{ total derivative }\label{VnlBRD=39}\ee

\no The equations of motion $\delta S^{D=3}_{NLBR}=0$ are satisfied by the flat space solution: $g_{\mu\nu}=\eta_{\mu\nu} \, ; \, \widehat{f}_{\mu\nu}=\widetilde{B}_{\mu\nu}=0$. Expanding about it, using
\be g_{\mu\nu}=\eta_{\mu\nu}+\frac{h_{\mu\nu}}{\sqrt{M_{P}}}\quad,\quad \tilde{f}_{\mu\nu}=\frac{f_{\mu\nu}}{\sqrt{M_{p}}}\quad,\quad \widetilde{B}_{\mu\nu}=\frac{b_{\mu\nu}}{\sqrt{M_{P}}}\label{VnlBRD=38}\ee

\no we recover the linearized $D=3$ mBR  model (\ref{BRD=35c}) at $M_P \to \infty $.
Now if we go beyond the linearized approximation and try to examine the decoupling limit as we have done in the bimetric model of the previous section
there will be an important difference. First, notice the expansion

\be \mathbb{G}_{\mu\nu}(\widetilde{B})=\mathbb{G}^{(L)}_{\mu\nu}(b)+\mathbb{G}^{(2)}_{\mu\nu}(b,h)+\mathbb{G}^{(3)}_{\mu\nu}(b,h^{2})+\ldots\label{VnlBRD=37}\ee
where $\mathbb{G}^{(n)}_{\mu\nu}(b)$ is of order $n$ in the fields. The tensor $\mathbb{G}^{(L)}_{\mu\nu}(b)$ coincides with the linearized Einstein tensor  for the field $b_{\mu\nu}$, i.e., $\mathbb{G}^{(L)}_{\mu\nu}(b)=G^{(L)}_{\mu\nu}(b)$. Thus, $\partial^{\mu}\mathbb{G}^{(L)}_{\mu\nu}(b)=0$, but  $\partial^{\mu}\mathbb{G}^{(2)}_{\mu\nu}(b,h)\neq0$. If we substitute the Stueckelberg covariant decomposition (\ref{LDbBR1}) in (\ref{VnlBRD=33}), after taking $ m\rightarrow0 $ and $M_{P}\rightarrow\infty $ while keeping $\Lambda_{5/2}=(\sqrt{M_{P}}m^{2})^{2/5}$ fixed, we obtain the following term at cubic order in the fields

\be \frac{\partial^{\mu}\partial^{\nu}\phi\,\mathbb{G}^{(2)}_{\mu\nu}(b,h)}{\Lambda^{5/2}_{5/2}}\label{cubic}\ee

\no which comes from  $\sqrt{-g}\widehat{f}^{\mu\nu}\mathbb{G}_{\mu\nu}(\widetilde{B})$. The term (\ref{cubic}) has more than two time derivatives and apparently introduces new degrees of freedom in the theory which will probably cause instabilities. The root of the problem is the non invariance of the integral of $\sqrt{-g}\widehat{f}^{\mu\nu}\mathbb{G}_{\mu\nu}(\widetilde{B})$ under
$\delta \widehat{f}^{\mu\nu}=\nabla^{\mu}\chi^{\nu} + \nabla^{\nu}\chi^{\mu}$ due to $\nabla^{\mu}\mathbb{G}_{\mu\nu}(\widetilde{B}) \ne 0$. Basically, the same problem appears in the general $D$-dimensional case for which we turn now.

A natural non linear completion of (\ref{mbr}) is given by


\be S^{D}_{BR}=\frac{1}{\kappa^{2}}\int{d^{D}x}\sqrt{-g}\Big\{R(g)-\frac{1}{2}\widehat{B}^{\mu\nu\alpha\beta}R_{\mu\nu\alpha\beta}(g)+\frac{1}{2}\widehat{B}^{\mu\nu\alpha\beta}\mathbb{R}_{\mu\nu\alpha\beta}(\widehat{f})
-\,\frac{m^{2}}{4}(\widehat{f}_{\mu\nu}\widehat{f}^{\mu\nu}-\widehat{f}^{2})\Big\}\nn\\ \label{VnlBRD3}\ee

\no where

\bea \mathbb{R}_{\mu\nu\alpha\beta}(\widehat{f})&=&\frac{1}{4}\Big(\nabla_{\nu}\nabla_{\alpha}\widehat{f}_{\mu\beta}-\nabla_{\nu}\nabla_{\beta}\widehat{f}_{\mu\alpha}-\nabla_{\mu}\nabla_{\alpha}\widehat{f}_{\nu\beta}+\nabla_{\mu}\nabla_{\beta}\widehat{f}_{\nu\alpha}\Big) \nn \\ &+&\frac{1}{4}\Big(\nabla_{\beta}\nabla_{\mu}\widehat{f}_{\alpha\nu}-\nabla_{\beta}\nabla_{\nu}\widehat{f}_{\mu\alpha}-\nabla_{\alpha}\nabla_{\mu}\widehat{f}_{\nu\beta}+\nabla_{\alpha}\nabla_{\nu}\widehat{f}_{\mu\beta}\Big)\label{VnlBRD5}\eea

\no Notice that

\be \sqrt{-g}\, \widehat{B}^{\mu\nu\alpha\beta}\mathbb{R}_{\mu\nu\alpha\beta}(\widehat{f}) = - 2\, \sqrt{-g}\,  \widehat{f}^{\mu\nu}\mathbb{S}_{\mu\nu}(\widehat{B}) + {\rm total \,\,\, \, derivative} \label{td2} \ee
\no where we have introduced the symmetric field:

\be \mathbb{S}_{\mu\nu}(\widehat{B}) \equiv \frac 12 \left(\nabla^{\alpha}\nabla^{\beta}\widehat{B}_{\alpha\mu\beta\nu} + \nabla^{\alpha}\nabla^{\beta}\widehat{B}_{\alpha\nu\beta\mu} \right) \label{smn} \ee

\no In $D=3$ the role of the symmetric  tensor $\mathbb{S}_{\mu\nu}(\widehat{B})$ is played by $ \mathbb{G}_{\mu\nu}(\widetilde{B})$. As in the $D=3$ case, the symmetric tensor is not conserved in general $\nabla^{\mu}\mathbb{S}_{\mu\nu}\ne 0$. Consequently, new degrees of freedom show up which may destroy stability. In particular, if we apply $\nabla^{\mu}$ on the equation of motion of $\widehat{f}^{\mu\nu}$ coming from (\ref{VnlBRD3}): $m^2(\widehat{f}_{\mu\nu}-g_{\mu\nu} \widehat{f})= -2 \, \mathbb{S}_{\mu\nu}(\widehat{B})$, because of the higher time derivatives on the right side we loose the curved space version of the flat space vector constraint $\p^{\mu}f_{\mu\nu}-\p_{\nu}f=0$ which is essential for a correct counting of degrees of freedom. The same problem can be seen from a different point of view, if we Gaussian integrate over $\widehat{f}^{\mu\nu}$ in (\ref{VnlBRD3}) we obtain a curved space version of the DTS model (\ref{BRD16}), equivalent to the Maxwell one on the flat space \cite{dts}, namely,

%

\be {\cal L}_{DTS}(\widehat{B}) = \frac {\sqrt{-g}}{m^2} \left( \mathbb{S}_{\mu\nu}(\widehat{B})\mathbb{S}^{\mu\nu}(\widehat{B})- \frac{\mathbb{S}^2(\widehat{B}
)}{D-1}\right) \label{lbbc}\ee

\no with $\mathbb{S}=g^{\mu\nu}\mathbb{S}_{\mu\nu}$. The reader can check that part of the curved space version of the local symmetries (\ref{gt2}) are lost. Indeed, on the flat space the restrictions (\ref{dtc}) on the tensor gauge parameter correspond to a symmetric transverse rank-2 tensor but  after replacing $\p^{\mu} \to \nabla^{\mu}$ and symmetrizing, the restriction is just a symmetric tensor, so we have extra $D$ restrictions on the curved space which means $D$ less symmetries. Moreover, the curved space version of the scalar symmetry: $\delta_{\pi}\widehat{B}_{\mu\nu\alpha\beta}= (g_{\mu\beta}g_{\nu\alpha}-g_{\mu\alpha}g_{\nu\beta})\pi $ is also broken since
 $\delta_{\pi}  [ \mathbb{S}_{\mu\nu}\mathbb{S}^{\mu\nu}- \mathbb{S}^2/(D-1)] = 2 \, \mathbb{S}^{\mu\nu}\nabla_{\mu}\nabla_{\nu}\pi $ and the integration by parts will not vanish. So the viability of the model (\ref{VnlBRD3}) is related with the  consistency of the  DTS model on curved spaces. In \cite{dts} a preliminary study of the curved space extension of gauge symmetries has been carried out without definite conclusion. Inspired by the procedure of \cite{phantom} in the definition of a nonlocal gravitational model, we recall \cite{deser,york} that any symmetric tensor can be decomposed  as $S_{\mu\nu} = S_{\mu\nu}^T + \nabla_{\mu}S_{\nu} + \nabla_{\nu} S_{\mu} $ where $\nabla^{\mu}S_{\mu\nu}^T=0$. So we might replace $\mathbb{S}_{\mu\nu}(B) \to \mathbb{S}_{\mu\nu}^T(B)= \mathbb{S}_{\mu\nu}(B)- \nabla_{\mu}S_{\nu} - \nabla_{\nu} S_{\mu} $ in (\ref{td2}) where the vector field $S_{\mu}$ must satisfy the vector condition $\nabla^{\alpha}\mathbb{S}_{\alpha\mu} - \nabla^2S_{\mu} - \nabla^{\alpha}\nabla_{\mu}S_{\alpha}=0$ which can be implemented by a vector Lagrange multiplier. This is currently under investigation.

\section{Conclusions}

The addition of mass terms usually breaks the gauge symmetry of massless theories which can be recovered by means of Stueckelberg fields. One exception to this rule is the $4D$ Cremmer-Scherck model \cite{cremmer} which describes massive spin $s=1$ particles while preserving the $U(1)$ symmetry, which can be generalized to the non Abelian case \cite{ft}, without Stueckelberg fields. In section II we have written the $D$- dimensional version of \cite{cremmer}, the massive BF model (mBF) model, in terms of some Lorentz non covariant gauge invariants. Likewise, we have written the $D$-dimensional generalization of the massive spin-2 model of \cite{ks}, which we call mBR  model, in terms of the corresponding  gauge invariants and argued that the mBR  model is very much, though not exactly, a spin-2 analogue of the mBF model. In both models a spin-s massless theory is coupled to a spin-(s-1) higher rank massless model by means of a gauge invariant mass term involving a spin-s curvature. We believe that there should be  higher spin ($s>2$) version of those massive models. In the $s=1$ case the mass term (BF term) couples the massless theories without breaking any gauge symmetry while in the $s=2$ case the BR term breaks the corresponding $U(1)$ symmetry of the higher rank spin-1 massless theory.

The mBR  model is a consistent description of massive free spin-2 particles.
The $D=3$ case is rather special because the rank-4 Riemann-like tensor $B_{\mu\nu\alpha\beta}$ is equivalent to a symmetric  rank-2 tensor via (\ref{BRD=36})
which may be interpreted as a second metric fluctuation about flat space which has inspired us to suggest a new class of bimetric models, see (\ref{bBR1}). Following \cite{yavin} we have gone beyond the linearized truncation and checked that the model is ghost free at leading order in the decoupling limit if the conditions (\ref{bBR21}) are satisfied. We can also add cosmological terms with independent cosmological constants for both metrics which leads to a rather general room to investigate the discrepancy between bulk and boundary unitarity of D=3 gravity in the context of $AdS_3/CFT_2$  duality \cite{more} in a purely metric formalism, differently from \cite{mmg,setare,emg}. Now we are investigating  a certain region in the parameters space of the model where we have found  AdS solutions with proportional metrics  $g_{\mu\nu} \propto \gamma_{\mu\nu} $ (in progress). In a future work we wish to study the stability of those solutions.

If, on the other hand, we stick to a single metric interpretation of a possible non linear completion of  the $D=3$ mBR  model, see (\ref{VnlBRD=33}), it turns out that the decoupling limit reveals the appearance of extra ghost-like degrees of freedom, see (\ref{cubic}). Basically the same problem goes thorough the $D\ge 4$ cases. We have shown that the consistency of the non linear model (\ref{VnlBRD3}) is tightly connected with the consistency of the higher rank massless spin-1 model of \cite{dts} on curved backgrounds. The key point is the non conservation of the symmetric tensor (\ref{smn}) on arbitrary backgrounds. We are currently investigating the replacement of $\mathbb{S}_{\mu\nu}(B) $ by  a transverse version $ \mathbb{S}_{\mu\nu}^T(B)$ as explained at the end of last section along the lines of \cite{phantom}.

\section{Acknowledgements}

 D.D. is partially supported by CNPq  (grant 313559/2021-0).

\vfill\eject

 \section{Appendix}

\subsection{Deser-Townsend-Siegel (DTS) model in $D$ dimensions}

The $D$-dimensional generalization of the $D=4$  DTS model \cite{dts} has been suggested in \cite{renato1},

\bea
\mathcal{L}_{DTS} =
\big(\partial_{\mu}\partial_{\nu}B^{\mu\alpha\nu\beta}\,\big)^{2}-\frac{1}{(D-1)}\big(\partial_{\alpha}\partial_{\beta}B^{\alpha\beta}\big)^{2}\label{ddts}
\eea
where $B^{\alpha\beta}=\eta_{\mu\nu}B^{\mu\alpha\nu\beta}$.


The DTS model is invariant under the transformations (\ref{gt2}) with the restrictions (\ref{dtc}). The number of gauge invariants built up from the field $B_{\mu\alpha\nu\beta}$ and its derivatives via (\ref{gt2}) is the number of independent components of $B_{\mu\alpha\nu\beta}$ minus the number of independent gauge parameters $(\Lambda_{\mu\alpha\nu\beta},\pi)$.
Since $\Lambda_{\mu\alpha\nu\beta}$ has the same index properties of $B_{\mu\alpha\nu\beta}$, the number of gauge invariants ($N_I$) is the number of restrictions (\ref{dtc}) minus 1 (due to $\pi$). The restrictions (\ref{dtc}) correspond to a symmetric transverse tensor, thus we have
$N_I=D(D-1)/2-1=(D+1)(D-2)/2$. In order to derive those invariants we start with the general decomposition (\ref{Bdec}) and decompose the gauge parameter in a similar way. After requiring (\ref{dtc})  we obtain
\bea
\Lambda_{[0i][0j]} &=& \Phi^{T}_{\,ij}+\partial_{i}\lambda^{T}_{j} + \partial_{j}\lambda^{T}_{i} \\
\Lambda_{[0i][jk]} &=& \Psi^{pT}_{i[jk]}+\partial_{j}\Theta^{pT}_{ik}-\partial_{k}\Theta^{pT}_{ij} \\
\Lambda_{[ij][kl]} &=& \Lambda^{T}_{[ij][kl]}
+\Big\{\partial_{i}\Omega^{T}_{j[kl]}+\partial_{k}\Omega^{T}_{l[ij]}
-\frac{\partial_{i}}{\nabla^{4}}\big[\partial_{k}\ddot\Phi^{T}_{\,j\,l}-\partial_{l}\ddot\Phi^{T}_{\,j\,k}\big]\nn\\
&&+\frac{\partial_{i}}{\nabla^{2}}\big[\partial_{k}(\dot\Theta^{pT}_{jl}+\dot\Theta^{pT}_{lj})-\partial_{l}(\dot\Theta^{pT}_{jk}+\dot\Theta^{pT}_{kj})\big]\Big\}
 \quad + \quad (i\leftrightarrow{j},k\leftrightarrow{l})\nn\\
\eea
with the constraint
\bea
\dot\lambda^{T}_{i}=\partial_{j}\Theta^{pT}_{ji} \label{constr-1}
\eea

%
%

\no From (\ref{gt2}) we have
 $\delta{B}_{[0i][0j]}=\Lambda_{[0i][0j]}-\delta_{ij}\pi$ which leads to
\bea
\pi &=& -\delta[2\nabla^{2}\rho] \label{0i0j-1}\\
\lambda^{T}_{i} &=& \delta{b}^{T}_{i} \label{0i0j-2} \\
\Phi^{T}_{ij} &=& \delta[B^{T}_{ij}-2(\partial_{i}\partial_{j}-\delta_{ij}\nabla^{2})\rho] \label{0i0j-3}
\eea
while from $\delta{B}_{[0i][jk]}=\Lambda_{[0i][jk]}$ we have
\bea
\Theta^{pT}_{ij} = \delta{C}^{pT}_{ij} \label{0ijk-1}\\
\Psi^{pT}_{i[jk]} = \delta{B}^{pT}_{i[jk]} \label{0ijk-2}
\eea
Applying $\partial_{i}$ in (\ref{0ijk-1}) and using the constraint (\ref{constr-1}) we find: $\dot\lambda^{T}_{j}=\delta[\partial_{i}{C}^{pT}_{ij}]$.
Combining this results with (\ref{0i0j-2}) we obtain  $\delta V_j^T =0$ where
\bea
 V_j^T &=& \dot{b}^{T}_{k}-\partial_{j}C^{pT}_{jk} \label{Vj}
\eea
On the other hand, applying $\partial_{j}\partial_{l}$ on
$\delta{B}_{[ij][kl]} = \Lambda_{[ij][kl]}+(\delta_{ik}\delta_{jl}-\delta_{il}\delta_{jk})\pi$
we have
\bea
\ddot\Phi^{T}_{ik}+\theta_{ij}\nabla^2\dot\Theta^{pT}_{jk}+\theta_{kj}\dot\Theta^{pT}_{ji} + \theta_{ik}\nabla^{2})\pi
=-\delta[\nabla^{4}W^{T}_{ik}] \label{ijkl}
\eea
Substituting (\ref{0i0j-1}), (\ref{0i0j-3}) and (\ref{0ijk-2}) in (\ref{ijkl}) we obtain $\delta \mathbb{W}_{ij}^T =0 $ where
\bea
\mathbb{W}_{ij}^T = \nabla^{4}W^{T}_{ij}+  \ddot{B}^{T}_{ij}+ \theta_{ik}\nabla^{2}\dot{C}^{pT}_{kj} +\theta_{jk}\nabla^{2}\dot{C}^{pT}_{ki}
-2\, \theta_{ij}\nabla^{2}\square\,\rho \label{Wij}
\eea

The invariant (\ref{Vj}) has $D-2$  components while in (\ref{Wij}) we have $(D-1)(D-2)/2$, they add up to a total of
 $(D+1)(D-2)/2$ gauge invariants as expected. After introducing the traceless and transverse tensor invariant $\omw_{ij} = \mathbb{W}_{ij}^T - \theta_{ij}\mathbb{W}/(D-2)$ where $\mathbb{W}=\delta^{ij}\mathbb{W}_{ij}^{T}$,
 the DTS Lagrangian density in $D$ dimensions can be written in a canonically simple way:

\be
\mathcal{L}_{DTS} = \frac 14 \left( \omw_{ij} \right)^2 + \frac {\mathbb{W}^2}{4\, (D-1)(D-2)} + \frac 12 V_j^T (-\nabla^2)\Box V_j^T \label{ldts}
\ee
So we have $D-2$ propagating massless degrees of freedom $ (V_j^T)$ as expected
for a massless spin-1 particle and $(D-1)(D-2)/2$
non propagating modes represented by $\mathbb{W}_{ij}^T$.

\end{document}